\documentclass[10pt,twocolumn]{article}

\usepackage[margin=.75in]{geometry}
\usepackage{authblk}
\usepackage[T1]{fontenc}
\usepackage{lmodern}
\usepackage[utf8]{inputenc}
\usepackage{amsmath,amssymb,booktabs,array}
\usepackage{graphicx,xcolor}
\usepackage[font=small,labelfont=bf]{caption}
\usepackage[numbers,sort&compress]{natbib}
\usepackage[hidelinks]{hyperref}
\usepackage{microtype}

\graphicspath{{figures/}}

\definecolor{paperpurple}{HTML}{7B5EA7}
\definecolor{paperteal}{HTML}{47968C}
\definecolor{papergray}{HTML}{6B727B}

\newcommand{\doi}[1]{\href{https://doi.org/#1}{doi:\nolinkurl{#1}}}

\newcommand{\MIN}{\mbox{\textsc{min}}}
\newcommand{\PRODUCT}{\mbox{\textsc{product}}}
\newcommand{\DEGREE}{\mbox{\textsc{degree}}}
\newcommand{\EIGEN}{\mbox{\textsc{eigenvector}}}

\newcommand{\variant}[2]{\mbox{#1-#2}}

\hypersetup{}

\title{A null model to detect coordinated online behavior \\ integrating evidence from multiple indicators}

\author[1,2]{Filipi Nascimento Silva}
\author[1]{Manita Pote}
\author[1]{Lake Yin}
\author[1]{David Axelrod}
\author[3]{Ashwin Balasubramanian}
\author[3]{Luca Luceri}
\author[1]{Filippo Menczer}
\author[1]{Alessandro Flammini}

\affil[1]{Observatory on Social Media, Indiana University, Bloomington, USA}
\affil[2]{Kellogg School of Management, Northwestern University, Evanston, USA}
\affil[3]{Information Sciences Institute, University of Southern California, Los Angeles, USA}

\begin{document}
\twocolumn[\begin{@twocolumnfalse}
\maketitle

\begin{abstract}
\noindent Coordinated online accounts can amplify messages and manipulate attention, but detecting such behavior is challenging due to confounding factors like account activity and item popularity. 
We propose a statistical null-model framework that assesses behavioral similarity between accounts by comparing observed coordination indicators against activity- and popularity-conditioned reference distributions. 
By combining evidence across multiple indicators and partners, our method can identify suspicious coordination links more accurately. 
Evaluations on nine labeled information-operation datasets involving over 152 thousand accounts show that aggregating p-values multiplicatively across indicators and partners yields improved detection performance. 
While results vary across information operations, this approach offers a generalizable and statistically principled tool for uncovering coordinated online behavior. 
\end{abstract}

\vspace{1em}\end{@twocolumnfalse}]

\section{Introduction}

Coordinated accounts can amplify messages, promote narratives, and manipulate online attention, posing significant challenges for the integrity of online discourse~\cite{CINELLI2022113819,Milzner18082026,Thiele26}. 
Detecting such activity requires examining relationships among accounts: participants may share links, amplify posts, act in synchrony, or use similar language without having unusual individual profiles~\cite{Francois23}. 
To capture these relationships, researchers often employ network-based methods that analyze shared behavioral traces~\cite{pacheco2021}. 
Applications include the detection of amplification networks~\cite{weber2021}, synchronized actions~\cite{magelinski2022}, and information operations~\cite{cima2024}.

Recent approaches have sought to improve detection by integrating multiple signals of information operations~\cite{luceri2024}, finding patterns in sharing speed and frequency~\cite{Yang2025}, representing both actions and content within a unified behavioral framework~\cite{nwala2023}, compressing such representations for detection~\cite{Loru26}, and monitoring how coordination patterns evolve over time~\cite{ariyarathne2026,tardelli2024temporal}. Other methods combine features of account profiles with evidence of shared narratives to identify coordinated activity~\cite{smith2025}.

Studies of election conversations find ordinary political participation alongside suspicious coordinated communities~\cite{nizzoli2021,tardelli2024multifaceted}, and comparisons with suitable controls can change conclusions about purported coordination~\cite{pante2025}. 
For example, we should not be surprised that two accounts reshare similar sets of content if that content is very popular. 
A key challenge is therefore to distinguish the similarity expected from item popularity and account activity from overlap that is not explained by these factors.

A second, related challenge is to effectively combine different indicators of coordinated behavior. For example, two accounts may exhibit moderately similar patterns of activity, such as using overlapping hashtags and resharing similar content, which individually may not appear suspicious. However, the combined evidence from multiple indicators could suggest non-independent behavior. Despite its importance, there has been little work on systematically integrating such evidence to assess the likelihood that two users are acting in coordination~\cite{iannucci2026detecting}. In particular, no general statistical framework currently exists to combine arbitrary coordination indicators.

\begin{figure*}
\centering
\includegraphics[width=\textwidth]{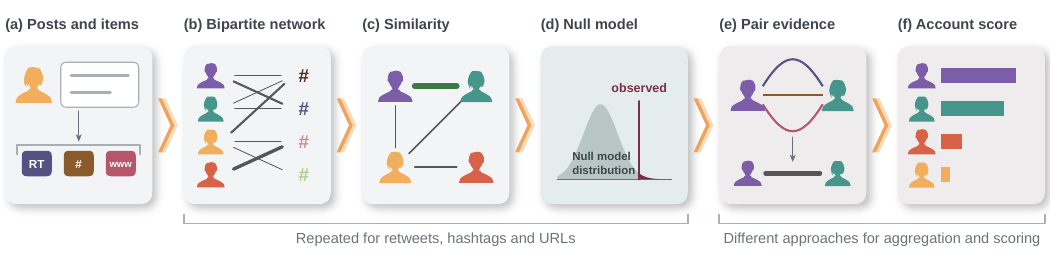}
\caption{\textbf{The proposed method pipeline in six steps.} (a)~Posts yield different types of indicators, for example retweeted posts, hashtags, and shared URLs. (b)~For each indicator type, accounts and items form an activity-weighted bipartite network. (c)~Account similarities are computed by projecting the bipartite network for each indicator into an account-account network. (d)~A null model that shuffles the bipartite network turns each similarity into a $p$-value. (e)~The $p$-values across different indicators are combined for each pair of accounts. (f)~The resulting edge weights are combined for account ranking. Figures~\ref{fig:data}--\ref{fig:aggregation} expand on each step.}
\label{fig:overview}
\end{figure*}

To address both of these gaps, we introduce a statistical null-model methodology designed to identify and combine suspicious behavioral similarity indicators within a general online coordination detection framework~\cite{pacheco2021, luceri2024}. 
Null models offer a principled way to evaluate whether observed patterns are statistically surprising. 
In network science, this idea has been applied to the detection of community structure~\cite{newman2006}, the rich-club phenomenon~\cite{colizza2006}, and network backbones~\cite{serrano2009}. 
In the context of online coordination, analysis often begins with a bipartite network connecting users to their online activities, which is then projected onto a user-user network to reveal pairs of users with unusually high behavioral similarity~\cite{pacheco2021}. 
Null models for such projected networks can account for differences in user activity and item popularity~\cite{tumminello2011,saracco2017,Neal24}, providing an appropriate reference for evaluating observed similarities.

We investigate the effects of different design choices within our proposed methodological pipeline, focusing on how evidence is aggregated across coordination indicators for each account pair, and subsequently across pairs to assign a single coordination score to each account. 
To evaluate these choices, we systematically compare eight variants of our approach using labeled information-operation datasets~\cite{seckin2025}. 
Although platform-provided labels do not specify which accounts are directly coordinated, they serve as an evaluation target at the account level, enabling us to assess how different strategies for indicator aggregation and account scoring influence the accuracy of account rankings.

\section{Methods}

The pipeline for our proposed method has six steps, as illustrated in Fig.~\ref{fig:overview}. 
Steps~b--d run separately for each indicator of coordinated behavior. Steps~e--f combine their output and are the object of our comparative evaluation.

\subsection{Null method}

\begin{figure*}
\centering
\includegraphics[width=\textwidth]{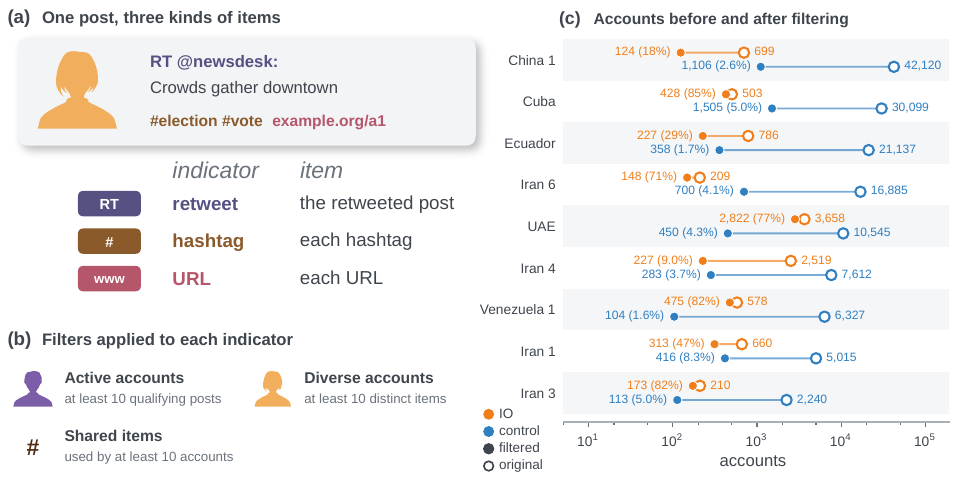}
\caption{\textbf{Data, indicators, and filters.} (a)~A post can contribute retweeted posts, hashtags, and URLs as three indicators. (b)~Three filters are applied before each indicator's network is built. (c)~Account counts before (open circles) and after filtering (filled circles) for the nine datasets, on a logarithmic scale. Lines connect the two counts for each account class. Filtered accounts pass at least one of the three indicator filters. Labels show raw counts and percentages retained after filtering, rounded to two significant digits. Orange and blue indicate information-operation (IO) and control accounts, respectively.}
\label{fig:data}
\end{figure*}

The method we present is broadly applicable to any social media platform where account activity indicators can be extracted from available data.

\subsubsection{Step 1: indicators and filters}
\label{sec:step1}

Each indicator captures a specific type of sharing behavior (Fig.~\ref{fig:data}(a)). 
In this study, we focus on three indicators: retweeted posts, hashtags, and URLs (web links). 
Hashtags and URLs are considered only when they appear in original tweets, not in retweets. 
A single post may retweet at most one post, but can include multiple hashtags and links. For the purposes of our analysis, each use of an item within a post is counted as one occurrence, even if the same account uses the item multiple times. 
While we focus on these three indicators, the method is sufficiently general to accommodate other types of coordination signals, such as keywords, topics, reshared accounts (distinct from reshared posts), timestamps, media, or API app clients.

Three filters are applied to each indicator (Fig.~\ref{fig:data}(b)): (i)~accounts must have at least 10 qualifying posts for that indicator; (ii)~items must be used by at least 10 distinct accounts; and (iii)~accounts must use at least 10 distinct items retained after applying (ii). 
These filters are applied sequentially, in the order listed, and are not repeated. 
An account is included in our analysis if it is retained after applying the three filters for at least one indicator (Fig.~\ref{fig:data}(c)).

\subsubsection{Step 2: account--item bipartite network}

\begin{figure*}[t]
\centering
\includegraphics[width=\textwidth]{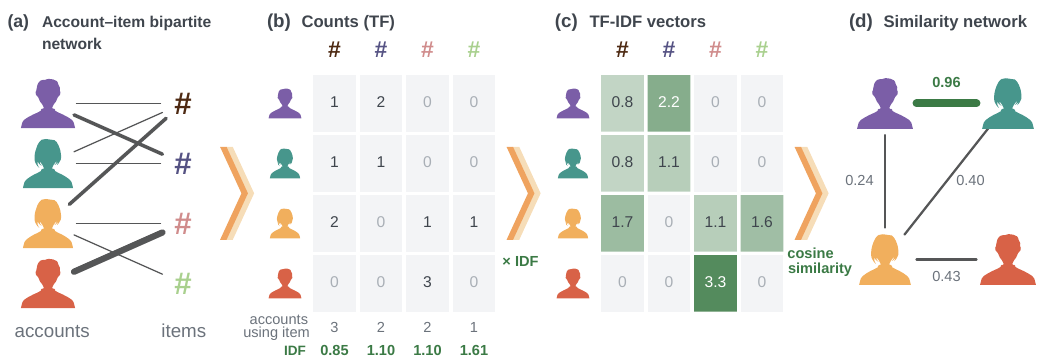}
\caption{\textbf{From a bipartite network to a similarity network.} (a)~Line width shows how often an account used an item. (b)~The same network as a table of counts (TF); the inverse document frequency (IDF) is larger for items used by fewer accounts. (c)~TF multiplied by IDF give one TF-IDF vector per account. (d)~In the projected account-account network, the cosine similarity between two TF-IDF vectors is the weight of the link between the two accounts. All numbers are computed from the example.}
\label{fig:similarity}
\end{figure*}

For any given indicator, let $c_{aj}$ be the number of times account $a$ used item $j$. 
These counts define a weighted bipartite network between accounts and items (Fig.~\ref{fig:similarity}(a)). 
A weight can equally be read as that many parallel edges, one per occurrence. 
The \emph{activity} of an account, $k_a=\sum_j c_{aj}$, is its number of occurrences, and the \emph{popularity} of an item is the number of occurrences it receives, $\sum_a c_{aj}$.

\subsubsection{Step 3: TF-IDF and cosine similarity}

Sharing a rare item is more informative than sharing a popular one. 
To capture this, we calculate the inverse document frequency of the item:
\begin{equation}
 \mathrm{idf}_j=\ln\left(1+\frac{N}{n_j}\right),
\end{equation}
where $N$ is the number of accounts and $n_j$ is the number of accounts that used item $j$ (Fig.~\ref{fig:similarity}(b)). 
Each count is multiplied by the inverse document frequency of the item to obtain an account's TF-IDF vector (Fig.~\ref{fig:similarity}(c)) with components: 
\begin{equation}
x_{aj} = c_{aj} \, \mathrm{idf}_j. 
\end{equation}
The similarity of two accounts is the cosine between their vectors (Fig.~\ref{fig:similarity}(d)): 
\begin{equation}
 s_{ab} = \frac{\mathbf{x}_a\cdot\mathbf{x}_b}{\lVert\mathbf{x}_a\rVert\,\lVert\mathbf{x}_b\rVert}.
\label{eq:cosine}
\end{equation}
The result is a weighted network among accounts. Pairs that share no item have similarity zero and no link.

\subsubsection{Step 4: null model and \texorpdfstring{$p$}{p}-values}

The same cosine similarity value can have different interpretations depending on the activity levels of the accounts involved. Highly active accounts are more likely to share many popular items by chance, whereas a similar degree of overlap between less active accounts may be more unusual. To account for this, we employ a null model that compares the observed similarity between two accounts to the expected distribution of similarities for that pair under random association.

\begin{figure*}
\centering
\includegraphics[width=\textwidth]{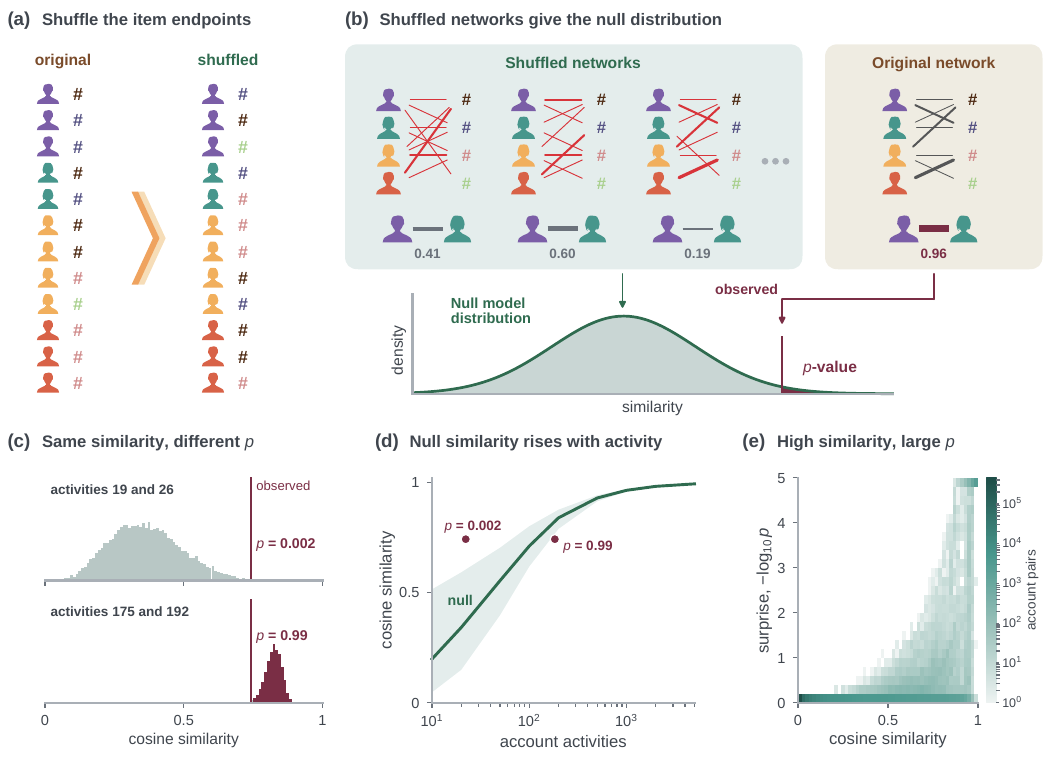}
\caption{\textbf{Activity conditions the interpretation of similarity.} (a)~Shuffling the item column of the occurrence list preserves account activity and item popularity exactly; pooled sampling approximates this reference in the implementation. (b)~Schematic null distribution for an activity pair. The upper tail gives the probability of a similarity at least as large as observed. (c--e)~Hashtag similarity results for the Cuba IO dataset. (c)~Two pairs with similarity near 0.74. The less active pair lies in the upper tail; the more active pair lies below its null median. Histograms use 20,000 independent draws; annotated $p$-values come from the saved 100,000-draw analysis. (d)~Null median and 5--95\% range for equally active pairs. The two example pairs are positioned at the geometric mean of their activities, so the curve provides a qualitative comparison. (e)~Linked account pairs: high similarity can coexist with large $p$-values (low surprise).}
\label{fig:nullmodel}
\end{figure*}

\textbf{Reference model.} We write the bipartite network as a list of (account, item) occurrences. 
Keeping the account column fixed and shuffling the item column preserves each account's activity and each item's popularity while randomizing their associations (Fig.~\ref{fig:nullmodel}(a)). 
This defines a configuration model for the bipartite network~\cite{saracco2017}. 
In our application, this approach is infeasible due to expensive IDF computations. 
Therefore, our implementation approximates this reference model through pooled sampling and fixed null IDF weights, as described below.

\textbf{Null similarities.} The null distribution of similarities for a pair of accounts $(a, b)$ depends on its two activities $(k_a, k_b)$. We therefore sample representative accounts for each distinct activity value and estimate a distribution for each activity pair using $B=100{,}000$ draws. Representative accounts receive items sampled from the occurrence list. Sampling is with replacement, which preserves item proportions in expectation rather than exact item counts. Null TF-IDF vectors use IDF values averaged over 1,000 fully shuffled networks. 

\textbf{$p$-values.} Let $\beta_{ab}$ be the number of null similarities at least as large as the observed similarity $s_{ab}$. For indicator $i$, the $p$-value of the pair is calculated as
\begin{equation}
 p^{(i)}_{ab}=\frac{1+\beta_{ab}}{1+B}
\label{eq:pvalue}
\end{equation}
as illustrated in Fig.~\ref{fig:nullmodel}(b). The add-one correction avoids zero estimates~\cite{phipson2010} and sets a minimum value of $1/(B+1) \approx 10^{-5}$. Pairs sharing no items receive $p=1$. 
A small $p$-value indicates that the observed similarity is suspiciously high under the implemented reference distribution. 

\textbf{Evidence from similarity vs.~$p$-values.}
To illustrate why similarity alone can be misleading in the identification of coordinated accounts, let us look at one of the datasets used in our evaluation, related to an information operation from Cuba (see Section~\ref{sec:methods:data}. 
Consider the IO network built from this dataset based on hashtag indicators, and in particular two pairs of accounts with comparable similarities but different activity levels (Fig.~\ref{fig:nullmodel}(c)). 
The less active pair lies in the upper tail ($p<0.01$), while the more active pair lies below its null median ($p>0.9$). 
These $p$-values demonstrate that the pair with lower activity is likely coordinated, while the pair with higher activity is not. 
Under the null model, the median similarity of pairs of equally active accounts rises from about 0.2 at ten occurrences to above 0.9 at 500 occurrences (Fig.~\ref{fig:nullmodel}(d)). 
Thus, substantial similarity can be explained by activity alone --- as well as by item popularity (not shown). 
This is why the two account pairs mentioned earlier have very different null distributions. 
A fixed similarity threshold would label as suspicious many pairs whose similarity is high but not surprising (Fig.~\ref{fig:nullmodel}(e)). 
Conversely, across a sample of roughly 800,000 linked pairs, fewer than 1\% have $p<0.01$ . 

\subsubsection{Step 5: aggregation to edges}

\begin{figure*}
\centering
\includegraphics[width=\textwidth]{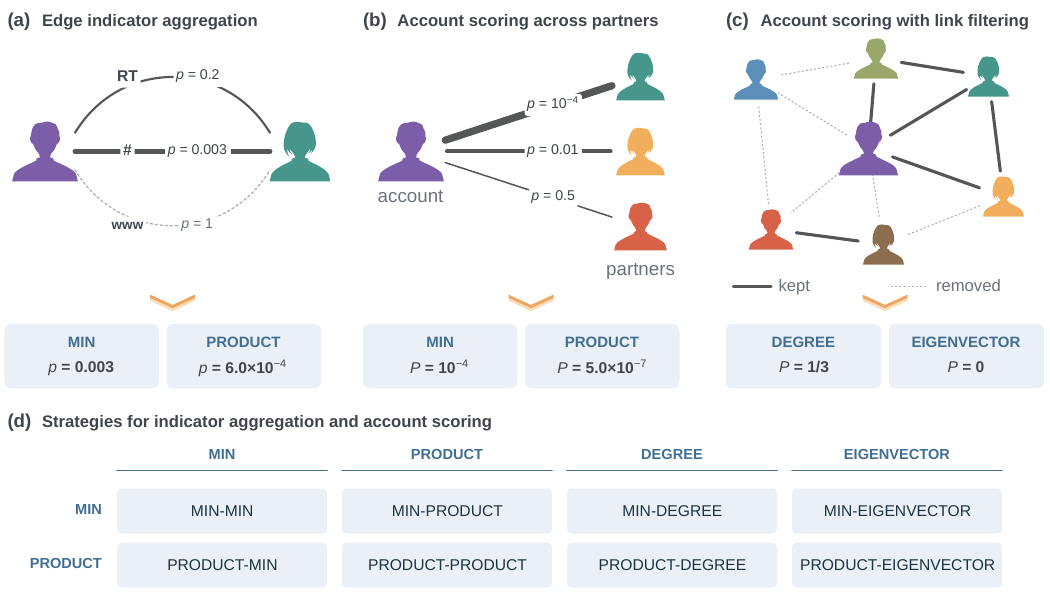}
\caption{\textbf{From indicator $p$-values to account rankings.} (a)~Indicator $p$-values are combined by their minimum or product to obtain an aggregated edge value. (b)~An account score $P$ can be obtained by aggregating across partner edge values via their minimum or product. (c)~Alternatively, we can filter out edges with $p>\alpha$ for some cutoff level $\alpha$. Accounts are then ranked by degree or eigenvector centrality in the resulting network. (d)~The eight variants: two edge indicator aggregations by four account scoring methods.}
\label{fig:aggregation}
\end{figure*}

After the previous step, a pair has one $p$-value for each of the indicators, $p^{(i)}_{ab}$. We compare two ways to combine evidence of coordination across indicators and obtain a combined edge value $p_{ab}$ (Fig.~\ref{fig:aggregation}a). \MIN{} keeps the smallest indicator $p$-value, emphasizing the strongest indicator: $p_{ab}=\min_i p^{(i)}_{ab}$. \PRODUCT{} multiplies the indicator $p$-values, accumulating evidence across indicators: $p_{ab}=\prod_{i} p^{(i)}_{ab}$. In both cases, smaller edge values indicate stronger evidence under the chosen aggregation. 

\subsubsection{Step 6: aggregation to accounts}

After combining evidence from different coordination indicators in the previous step, we obtain a weighted network where two accounts $a,b$ may be linked by an edge with weight $p_{ab}$. 
The final step in our pipeline is to leverage this network to assign scores to individual accounts, which can be used to rank accounts by suspiciousness. 
We present four methods to compute account scores (Fig.~\ref{fig:aggregation}(b,c)).

Let $\mathcal{N}(a)$ be the set of neighbors of account $a$ in the network (i.e., the accounts with which $a$ shares at least one item in at least one indicator). 
\MIN{} assigns the account value $P_a = \min_{b\in\mathcal{N}(a)} p_{ab}$, considering the most suspicious coordination partner. \PRODUCT{} assigns $P_a = \prod_{b\in\mathcal{N}(a)} p_{ab}$, combining evidence from all potential partners. 
Accounts without partners receive $P=1$ under either rule.

Two additional methods require an extra network pruning step, where edges with $p_{ab}>\alpha$ are removed and the remaining edges form an unweighted network $G_{\alpha}$. Here the parameter $\alpha$ is a cutoff threshold. \DEGREE{} scores an account by the reciprocal of its number of links (degree) in $G_{\alpha}$, and \EIGEN{} by one minus its normalized eigenvector centrality, according to the idea that the network may provide useful coordination signals beyond the immediate local neighborhood of a node \cite{luceri2024}. 
Accounts without retained links are assigned $P=1$.

In all of these methods, accounts with smaller $P$ values are considered more suspicious. 

\subsection{Data}
\label{sec:methods:data}

\begin{table}
\centering
\setlength{\tabcolsep}{3.5pt}
\caption{\textbf{Account populations and post counts.} 
Accounts ``with evidence'' are those passing at least one indicator filter. 
}
\begin{tabular}{@{}lrrrrr@{}}
\toprule
 & & \multicolumn{2}{c}{Accounts} & \multicolumn{2}{c}{With evidence} \\
\cmidrule(lr){3-4}\cmidrule(l){5-6}
Dataset & Posts & IO & Control & IO & Control \\
\midrule
China 1 & 3,670,754 & 699 & 42,120 & 124 & 1,106 \\
Cuba & 6,155,331 & 503 & 30,099 & 428 & 1,505 \\
Ecuador & 1,470,529 & 786 & 21,137 & 227 & 358 \\
Iran 1 & 1,322,789 & 660 & 5,015 & 313 & 416 \\
Iran 3 & 2,054,438 & 210 & 2,240 & 173 & 113 \\
Iran 4 & 561,585 & 2,519 & 7,612 & 227 & 283 \\
Iran 6 & 1,289,812 & 209 & 16,885 & 148 & 700 \\
UAE & 1,707,792 & 3,658 & 10,545 & 2,822 & 450 \\
Venezuela 1 & 1,246,290 & 578 & 6,327 & 475 & 104 \\
\bottomrule
\end{tabular}
\label{tab:populations}
\end{table}

Our evaluation leverages datasets in the information-operation collection described by Seckin et al.~\cite{seckin2025}. 
Each dataset pairs accounts attributed by a platform to an information operation (IO), and therefore assumed to be coordinated, with control accounts selected through related topics and dates. 
We extract three indicators: retweeted posts, hashtags, and URLs. 
We require a dataset to have at least 50 uniquely labeled IO accounts and 50 controls after filtering for the three indicators. 
Nine datasets meet this inclusion rule, spanning six state actors and comprising about 19 million posts and 152 thousand accounts, including about 10,000 IO accounts before filtering. 
Dataset summary statistics are reported in Table~\ref{tab:populations}.

\subsection{Evaluation}

We consider eight variants within our framework (Fig.~\ref{fig:aggregation}(d)). A specific method is named by its edge aggregation and account scoring method, edge level first. For example, \variant{\PRODUCT}{\MIN} multiplies indicator $p$-values for each pair and then ranks an account by its strongest partner, while \variant{\PRODUCT}{\PRODUCT} multiplies the resulting edge values across all partners. 
The eight variants in our comparison result from two edge aggregations combined with four account scoring methods.
Indicators and partners are not independent, and no multiple-testing correction is applied, so neither $p_{ab}$ nor $P_a$ is a calibrated $p$-value.

Each variant ranks the accounts in a dataset by their $P$ values, with lower values indicating higher suspicion. 
The evaluation includes only those accounts that pass at least one of the three indicator filters. 
We assess the rankings against the IO labels using the area under the ROC curve (AUC), average precision (AP), and the precision among the top 100 ranked accounts (P@100). 
Precision-based measures are particularly informative when class distributions are imbalanced~\cite{saito2015}. 
Ties are handled exactly: AP is calculated at distinct score values, and P@100 is averaged over all accounts tied at the cutoff. 
The reported mean values are averaged across the nine eligible datasets, each weighted equally.

\section{Results}

\begin{figure}[t]
\centering
\includegraphics[width=\columnwidth]{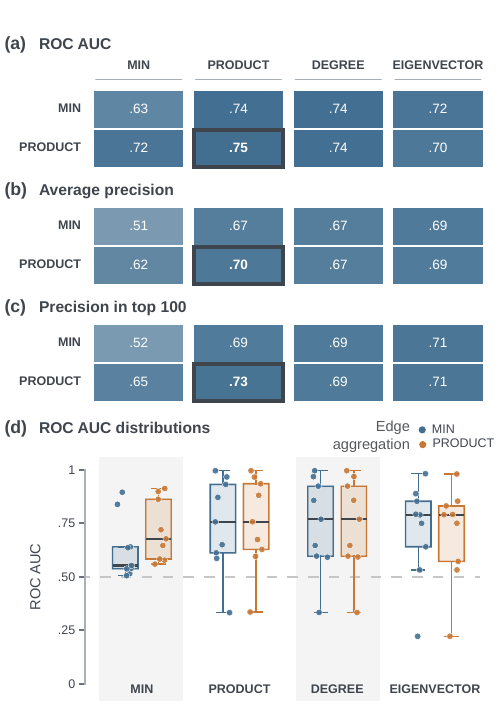}
\caption{\textbf{Performance of the eight variants.} Each variant uses a particular combination of edge indicator aggregation and account scoring methods. \DEGREE{} and \EIGEN{} use $\alpha=0.05$. Values are means over nine datasets. (a)~ROC AUC. (b)~Average precision. (c)~Precision among the top 100 accounts. Darker shading indicates higher scores on a shared 0--1 scale. The outlined cell is the best for each measure. (d)~Dataset-level ROC AUC, grouped by account method, with \MIN{} (blue) and \PRODUCT{} (orange) edge indicator aggregation. The nine dataset values are shown as circles.}
\label{fig:variants}
\end{figure}

The comparison covers the eight variants using retweeted post, hashtag, and URL indicators, with $\alpha=0.05$ for \DEGREE{} and \EIGEN{} scoring (Fig.~\ref{fig:variants}). Aggregating partner edge values with \PRODUCT{} yields higher mean performance than retaining only the strongest partner with \MIN{}, under both edge aggregation rules (Fig.~\ref{fig:variants}(a--c)). The highest mean values across all three measures are obtained with \variant{\PRODUCT}{\PRODUCT}, whose mean AUC is about 0.75. The \DEGREE{} variants achieve similar mean AUC, with slightly lower average precision and P@100. Changing the edge aggregation from \MIN{} to \PRODUCT{} has its clearest effect when the account method is \MIN{}, whereas the \DEGREE{} results remain closely similar under both edge aggregation methods.

Significant variation across datasets is visible in the box-plot representation of the AUC distributions (Fig.~\ref{fig:variants}(d)). The \PRODUCT{} and \DEGREE{} account methods have similar distributions, while \MIN{} and \EIGEN{} show different patterns. The AUC ranges nevertheless broadly overlap among most variants, with \variant{\MIN}{\MIN} resulting in a more restricted range.

\begin{figure}
\centering
\includegraphics[width=\columnwidth]{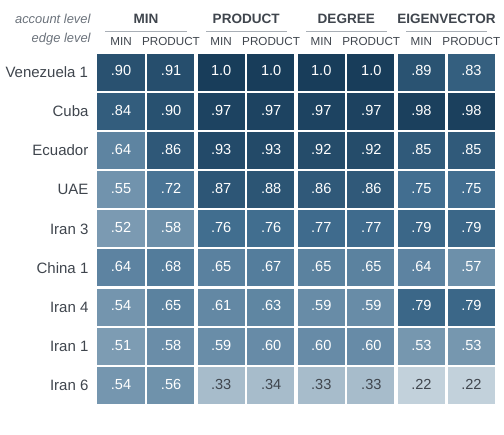}
\caption{\textbf{Dataset-level ROC AUC for the eight variants.} Columns are grouped by account scoring methods, each with \MIN{} and \PRODUCT{} edge indicator aggregation. \DEGREE{} and \EIGEN{} retain links using $\alpha=0.05$. Darker shading indicates higher AUC.}
\label{fig:datasets_auc_main}
\end{figure}

Not all information operations present equally challenging detection cases, as illustrated by the dataset-level results in Fig.~\ref{fig:datasets_auc_main}. 
All eight variants achieve high AUC for the Cuba and Venezuela~1 information operations, suggesting comparatively easier detection of coordinated accounts for the indicators and methods considered. In contrast, Iran~6 appears more challenging, with six variants yielding AUC below 0.5. The variant attaining the highest AUC ia also not consistent across datasets.

\begin{figure}[t]
\centering
\includegraphics[width=\columnwidth]{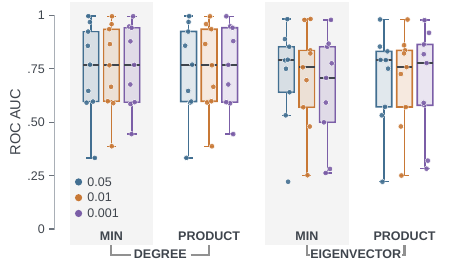}
\caption{\textbf{Sensitivity of ROC AUC to the edge value threshold $\alpha$.}
The four groups combine \DEGREE{} and \EIGEN{} account scoring methods with \MIN{} and \PRODUCT{} edge aggregation. Within each group, boxes show $\alpha=0.05$ (blue), $0.01$ (orange), and $0.001$ (purple), from left to right. Edges are retained when $p_{ab}<\alpha$. Each box summarizes AUC value distribution across the nine datasets (also shown as points).}
\label{fig:auc_thresholds}
\end{figure}

The effects of using stricter edge-value thresholds ($\alpha$) are weak and depend on the account scoring metric  (Fig.~\ref{fig:auc_thresholds}). For \DEGREE{}, the AUC distributions remain closely similar across the three thresholds. \EIGEN{} is slightly more sensitive, particularly with \MIN{} edge aggregation, for which stricter cutoffs shift the distribution toward lower AUC values. 

\section{Conclusion}

Our results illustrate how similar online sharing behaviors can be assessed for statistical anomalies using a reference model and statistical rules to combine evidence from multiple coordination indicators.
An activity- and popularity-conditioned null model can assign very different evidence to account pairs with comparable behavioral similarity. 
In this way, one can identify pairs of accounts whose similarity is statistically unlikely under a model conditioned on account activity and item popularity.

Multiplying $p$-values across retweet, hashtag, and URL indicators, and subsequently across partner accounts, yields modestly higher mean AUC, AP, and P@100 for detecting coordinated accounts. 
However, substantial variation across datasets precludes strong conclusions about a single preferred variant of the proposed method.

Several limitations affect the interpretation of the null-model evidence. 
Actions are aggregated without considering temporal windows.
Item popularity is estimated from data that may include coordinated accounts, potentially biasing the reference distribution. 
URLs are compared after query-string removal, with no further canonicalization. 
Replacement sampling preserves item proportions only on average, and the use of averaged null IDF weights introduces additional approximation. 
Finally, the null model does not account for topics, communities, or temporal dynamics, so observed overlap may sometimes reflect shared interests rather than coordination. 

Future work could examine the sensitivity of our proposed methods to these limitations and compare their performance with other coordination detection approaches in the literature. 
The null model can also be applied to improve existing methods. 
For example, similarity networks weighted by $p$-values could be used in conjunction with community detection algorithms~\cite{iannucci2026detecting} to discover loosely coordinated groups. 

Progress in this area is currently hampered by the lack of transparency from most social media platforms, which makes it extremely difficult to obtain empirical data on past and ongoing information operations~\cite{seckin2025}. 
Increased transparency from platforms would greatly facilitate research aimed at developing effective strategies to mitigate the harms of online manipulation.




\section*{Acknowledgments}

This work was supported in part by DARPA (contract HR001121C0169) and the Knight Foundation. 
We utilized Indiana University Jetstream2 CPU through allocation CIS230183 from the Advanced Cyber-infrastructure Coordination Ecosystem: Services \& Support (ACCESS) program, which is supported by National Science Foundation grants 2138259, 2138286, 2138307, 2137603, and 2138296. 
Any opinions, findings, and conclusions or recommendations expressed in this material are those of the authors and do not necessarily reflect the views of the funders.  

\begingroup\small\raggedright
\bibliographystyle{unsrtnat}
\bibliography{coord_null}
\endgroup
\end{document}